\documentstyle[aps, prl, twocolumn, graphics, epsfig,
floats]{revtex}

\newcommand{\beq}{\begin{equation}}
\newcommand{\eeq}{\end{equation}}
\newcommand{\beqa}{\begin{eqnarray}}
\newcommand{\eeqa}{\end{eqnarray}}

\def\opone{\leavevmode\hbox{\small1\kern-3.8pt\normalsize1}}

\begin{document}

\title{Hidden-variable theorems for real experiments}

\author{Christoph Simon$^{1,2}$, \v{C}aslav Brukner$^1$, and Anton Zeilinger$^1$}

\address {$^1$
Institut f\"ur Experimentalphysik, Universit\"at Wien, Boltzmanngasse 5, A-1090 Wien,  Austria\\$^2$ Centre for
Quantum Computation, Clarendon Laboratory, University of Oxford, Parks Road, Oxford OX1 3PU, United Kingdom}

\date{\today}
\maketitle

\begin{abstract}

It has recently been questioned whether the Kochen-Specker theorem is relevant to real experiments, which by
necessity only have finite precision. We give an affirmative answer to this question by showing how to derive
hidden-variable theorems that apply to real experiments, so that non-contextual hidden variables can indeed be
experimentally disproved. The essential point is that for the derivation of hidden-variable theorems one does not
have to know which observables are really measured by the apparatus. Predictions can be derived for observables
that are defined in an entirely operational way.

\end{abstract}

\vskip 1pc

In general quantum mechanics only makes probabilistic predictions for individual events. The question whether one
can go beyond quantum mechanics in this respect has been a subject of debate and research since the early days of
the theory. There are famous theorems placing restrictions on possible hidden-variable theories reproducing the
results of quantum mechanics. Bell's theorem \cite{bell} excludes local hidden-variables. The Kochen-Specker
\cite{ks} theorem excludes non-contextual hidden variables. In local hidden-variable theories the pre-determined
results for a given measurement are independent of which measurements are performed at space-like separation. In
non-contextual hidden-variable theories the pre-determined results are independent of {\em any} measurements that
are performed jointly.

The present work is concerned with the derivation of hidden-variable theorems that apply to real experiments. It
was motivated by the thought--provoking work of Meyer \cite{meyer}, who claimed that the Kochen-Specker theorem was
``nullified'' for real experiments because of the unavoidably finite measurement precision, i.e. because the
observer does not have full control over his experimental setup. However, our approach is very general and applies
in a much broader context. It allows one to deal with all possible experimental imperfections, and it is not
restricted to the specific class of non-contextual hidden variables. In particular, theorems on local hidden
variables can be derived in the same way.

Let us begin our discussion with some remarks on experimental tests of hidden variables in general. Hidden-variable
theorems provide us with predictions made by certain classes of hidden-variable theories which can be tested
experimentally and which differ from the quantum mechanical predictions. For local hidden variables, Bell's
inequalities can be derived without making any reference to quantum mechanics. Then one can check whether they are
fulfilled by the quantum mechanical predictions, and by experiment. For non-contextual hidden variables, one can
use the Kochen-Specker argument to derive specific quantitative predictions for the results of experiments, which
can then be compared to quantum mechanics and to nature. For experimental proposals and Kochen-Specker type
experiments that were performed in this spirit see \cite{ksexp}.

It is sometimes argued that the Kochen-Specker theorem just makes a conceptual point about the formal structure of
quantum mechanics and that actually performing experimental tests for non-contextual hidden variables is
unnecessary, since quantum mechanics is a very well tested theory. Irrespective of this discussion, we feel that
the question whether such tests could be performed {\em in principle} is interesting because it concerns the
falsifiability of a rather simple and fundamental concept.

In the derivation of hidden-variable theorems frequently no direct reference to experiments is made, or the
experimental situation is treated in an idealized way. For example, in the original derivation of the
Kochen-Specker theorem it was just shown that non-contextual hidden variables are incompatible with quantum
mechanics, without making the experimental predictions of non-contextual hidden variables explicit, while in the
original derivation of Bell's theorem perfect correlations and perfect detection efficiency were assumed.

If one wants to consider an experimental test for a certain class of hidden-variable theories, one has to give up
these idealized assumptions and derive hidden-variable theorems that apply to the true experimental situation. For
local hidden variables, the case of non-perfect correlations and detection efficiency was analyzed soon after
Bell's original derivation, starting with the work of Clauser and Horne \cite{clauser}. The same kind of analysis
is possible for non-contextual hidden variables.

However, there is one more idealization made in the usual derivations of hidden-variable theorems which, to our
knowledge, has never been discussed in an explicit way, namely the accuracy with which the experimental setup can
be maintained (e.g. how well the Stern-Gerlach magnets in a spin measurement can be aligned and kept stable). This
question has become particularly important for the Kochen-Specker theorem in view of recent claims by Meyer
\cite{meyer} that this theorem is no longer applicable when the measurements have only finite precision.

In its original form, the Kochen-Specker theorem states that is is impossible to assign values to observables
corresponding to all directions on the sphere subject to a constraint for triads of orthogonal directions. The
observables under consideration are the squares of the spin components of a spin-1 particle along the respective
direction, and the constraint is given by the total spin.

Meyer's claim was based on the fact that it is possible to assign values compatible with the constraint to all
rational directions on the sphere, which constitute a dense subset of all directions \cite{kent}. He argued that,
since measurements with finite precision cannot discriminate a dense subset from its closure, this implies that
non-contextual hidden variables cannot be excluded by any real Kochen-Specker type experiment. However, he did not
construct an explicit non-contextual hidden-variable model for real experiments with finite precision
\cite{clifton}.

At first sight, it does not seem possible to refute Meyer's claim on the basis of existing theorems, for the
following reason. An essential feature of all hidden-variable theorems known to us is that some observables have to
appear in different experimental contexts: an observable $A$ has to be measured together with another observable
$B$ on a sample of systems, and also together with a third observable $C$ on another sample of systems from the
same source, where both $B$ and $C$ commute with $A$, but they do not commute with each other and thus cannot be
measured jointly. Note that even for tests of non-contextuality it is not necessary to perform several incompatible
measurements on {\it the same} system, which would of course be impossible \cite{samples}.

For example, in the original Kochen-Specker situation, one can only arrive at a contradiction by considering
several triads of directions that have at least some directions in common. For Kochen-Specker experiments this
implies that, at least for some directions, the observables corresponding to these directions have to appear in
different triads. When the finite precision of real experiments is taken into account, it seems impossible to
ascertain that the {\it same} observable is really measured more than once in different experimental contexts. Thus
the usual derivations of hidden-variable theorems seem to run into problems.

Does this mean that it is impossible in principle to rule out non-contextuality experimentally? Could even the
experimental results on local hidden variables be in danger for the same reasons? We are going to show that the
answer is no. There are predictions of non-contextual hidden variables, which can be derived within a framework
that is sufficiently general to apply to real experiments, and which can therefore also be tested by real
experiments with sufficiently high, but finite, precision.

In order to achieve this, it is important to realize that predictions for classes of hidden-variable theories can
be derived without reference to quantum mechanics. In such an approach, the observables playing a role in the
hidden-variable predictions are not defined via the quantum-mechanical formalism, but in an {\it operational} way.

For concreteness, imagine that an observer wants to perform a measurement of the spin square of a spin-1 particle
along a certain direction $\vec{n}$. There will be an experimental procedure for trying to do this as accurately as
possible. We will refer to this procedure by saying that he sets the ''control switch'' of his apparatus to the
position $\vec{n}$. In all experiments that we will discuss only a finite number of different switch positions is
required. By definition different switch positions are clearly distinguishable for the observer, and the switch
position is all he knows about. Therefore, in an operational sense the measured physical observable is entirely
defined by the switch position. From the above definition it is clear that the same switch position can be chosen
again and again in the course of an experiment (while of course the system measured will always be different, cf.
\cite{samples}).

In general one has to allow for the possibility that the switch position $\vec{n}$ does not uniquely determine the
physical state of the measuring apparatus, i.e. there may be degrees of freedom of the apparatus over which the
observer does not have full control but which may influence the result of any given measurement. In the context of
deriving hidden-variable theorems, this possibility can be accomodated in a very simple and general way. Following
the philosophy of deterministic hidden variable theories \cite{stochastic}, there must also be some (in general
hidden, i.e. unknown) variables determining the behaviour of the apparatus, and one has to assume that the result
of any measurement will be determined by the hidden variables of the system and by those of the apparatus together.

Notice that in such an approach as described in the two preceding paragraphs it does not matter which observable is
``really'' measured by the apparatus and to what precision. One just derives general predictions for the behaviour
of system and apparatus together, provided that certain switch positions are chosen. These predictions only depend
on the properties of the class of hidden-variable theories considered. The question of the correct {\em
quantum-mechanical} description of the non-ideal measurement considered arises only when, as a next step, one wants
to obtain the quantum-mechanical predictions for the given situation.

Following the method described above, we will now show how non-contextual hidden variables can be tested and thus
potentially excluded by real experiments. We will consider the context of the original Kochen-Specker argument,
i.e. exactly the case considered by Meyer.

In the original Kochen-Specker situation one considers a spin-1 particle. In the ideal case of perfect precision,
the relevant quantum-mechanical observables are the squares of the spin components, denoted by
$\hat{S}^2_{\vec{n}}$ for arbitrary directions $\vec{n}$. For a spin-1 particle one has
\begin{equation}
\hat{S}^2_{\vec{n}_1}+\hat{S}^2_{\vec{n}_2}+\hat{S}^2_{\vec{n}_3}=2
\label{sum}
\end{equation}
for every triad of orthogonal directions $\{\vec{n}_1,\vec{n}_2,\vec{n}_3\}$. As the possible results for every
$\hat{S}^2_{\vec{n}_i}$ are 0 or 1, this implies that in the ideal case for every measurement of three orthogonal
spin squares two of the results will be equal to one, and one of them will be equal to zero.

Let us emphasize that in our approach, in the derivation of the hidden-variable predictions, the observables are
defined operationally by the switch positions, i.e. by the best effort and knowledge of the experimenter, and
cannot be identified with the quantum-mechanical observables. Of course, for a specific experiment, there should be
some approximate correspondence in order to ensure that the quantum-mechanical predictions will be sufficiently
close to the ideal case so that they are still in conflict with the relevant hidden-variable predictions. In the
following the symbol $S^2_{\vec{n}}$ (without the hat) will denote the operational observable defined by the switch
position $\vec{n}$, and the term direction will be used as a synonym for switch position.

In a deterministic hidden variable theory (cf. \cite{stochastic}) one assumes that for every individual particle
the result of the measurement of any observable $S^2_{\vec{n}}$ is predetermined by hidden variables. In {\it
non-contextual} hidden variable theories it is furthermore assumed that this predetermined result does not depend
on the ''context'' of the measurement, i.e. on which other observables are measured together with $S^2_{\vec{n}}$,
but only on the switch position $\vec{n}$ and the hidden variables \cite{context}.

In general the result may depend both on the hidden variables of the system and of the apparatus. Let us denote the
hidden variables of the system by $\lambda$ and those of the apparatus by $\mu$.  As explained above, the
philosophy of non-contextual hidden variables implies the existence of a function $S^2_{\vec{n}}(\lambda,\mu)$
taking values 0 and 1 which describes the result of a measurement with switch position $\vec{n}$ on a system
characterized by $\lambda$ with an apparatus characterized by $\mu$. For fixed $\lambda$ and $\mu$ this function
therefore assigns a value 0 or 1 to the switch position $\vec{n}$ \cite{clifton}.

A Kochen-Specker experiment can now be performed by testing the validity of Eq. (\ref{sum}) (without hats) for a
judiciously chosen set of triads of directions. Therefore the apparatus is required to have three switches where
the three directions of a given triad can be chosen. Because the switch positions do not correspond to the ideal
quantum mechanical observables the sum of the three results will not always be equal to 2. Nevertheless a
contradiction between non-contextuality and quantum mechanics can be obtained in the following way.

From the Kochen-Specker theorem it follows that there are finite sets of triads for which no value assignment
consistent with Eq. (\ref{sum}) (again without the hats) is possible \cite{ks,peres}. Let us choose such a
Kochen-Specker set of triads
\begin{equation}
\left\{ \{\vec{n}_1,\vec{n}_2,\vec{n}_3\},
\{\vec{n}_1,\vec{n}_4,\vec{n}_5\}, ..., \right\}. \label{ksset}
\end{equation}
Let us emphasize that at least some of the switch positions $\vec{n}_i$ have to appear in several of the triads;
clearly otherwise there could be no inconsistency. Let us denote the number of triads in the Kochen-Specker set
(\ref{ksset}) by $N$. The set is constructed in such a way that for any fixed values of $\lambda$ and $\mu$ the
equation
\begin{equation}
S^2_{\vec{n}_i}(\lambda,\mu)+S^2_{\vec{n}_j}(\lambda,\mu)+S^2_{\vec{n}_k}(\lambda,\mu)=2
\label{sumlambda}
\end{equation}
must be violated for at least one out of the $N$ triads $\{\vec{n}_i,\vec{n}_j,\vec{n}_k\}$.

Suppose that one can establish experimentally that for all triads in the  Kochen-Specker set the sum of the results
is equal to 2 in a fraction greater than $1-\epsilon$ of all cases. For the hidden variables this implies that Eq.
(\ref{sumlambda}) must hold for a fraction $1-\epsilon$ of all $(\lambda,\mu)$, for all triads in the set. But for
sufficiently small $\epsilon$ this implies that there would have to be pairs $(\lambda,\mu)$, for which Eq.
(\ref{sumlambda}) holds for all triads in the set, which is impossible according to the Kochen-Specker argument.

To determine the required value for $\epsilon$, it is convenient to use a set-theoretic language. Let us denote the
set of all pairs of hidden variables $(\lambda,\mu)$ by $\Lambda$. Furthermore let us denote the subset of hidden
variables for which the sum of spin squares is equal to $2$ for the $k$-th triad by $\Lambda_k$. The value of
$\epsilon$ has to be sufficiently small such that the intersection of all the $\Lambda_k$ cannot be empty. If we
define the measure (i.e. the size in terms of probability) of $\Lambda$ to be $1$, then according to our
assumptions all the $\Lambda_k$ have measures larger than $1-\epsilon$, which implies that the measure of the
intersection of all the $\Lambda_k$ is larger than $1- N
\epsilon$. This follows from the fact that the complement of each
$\Lambda_k$ is smaller than $\epsilon$, such that the size of the union of these complements, which is the
complement of the intersection of all the $\Lambda_k$, cannot be larger than $N
\epsilon$.

Thus non-contextuality is experimentally disproved as soon as $\epsilon$ is smaller than $1/N$, because then there
would have to be hidden variables which lead to a sum of spin squares equal to $2$ for all the triads, which is
impossible because of the structure of the Kochen-Specker set. Note that $\epsilon$ describes all the imperfections
of a real experiment including finite precision but also e.g. imperfect state preparation and non-unit detection
efficiency. The value of $N$ and therefore the bound on $\epsilon$ depends on the particular Kochen-Specker set
used \cite{ks,peres}.

As noted above, an inevitable requirement for the contradiction to be obtained is the fact that the function
$S^2_{\vec{n_1}}(\lambda,\mu)$, or in general functions corresponding to at least some switch positions, appear in
more than one out of the $N$ triads. This appearance of the same function in different lines of the mathematical
proof, corresponding to different experimental contexts, is possible in spite of finite experimental precision only
because we defined our observables operationally via the switch positions.

We have shown how non-contextual hidden-variable theories can be disproved by real experiments. This shows that
Meyer's coloring of a dense subset of the sphere cannot be used to construct non-contextual hidden-variable
theories according to the above definition. The values assigned to specific switch positions would have to depend
on the context, i.e. on the other switch positions chosen simultaneously by the experimenter. This form of
contextuality is a feature of existing explicit models based on Meyer's idea \cite{clifton}. In view of our
results, we would assert that the Kochen-Specker theorem is not ''nullified'' by finite measurement precision. Let
us note that arguments in favor of this conclusion were given in \cite{mermin2,apple}. Our suggestion how to
perform a Kochen-Specker experiment was inspired by some of Mermin's remarks in \cite{mermin2}.

Let us emphasize that using the method of the present paper one can also show that {\em local} hidden variables can
be disproved in real experiments, e.g. using the GHZ \cite{ghz} form of Bell's theorem which is also based on sets
of propositions that cannot be consistently satisfied by a class of hidden-variable theories (local, in this case).
Inequalities corresponding to our bound on $\epsilon$ can be derived and compared to the experimental data
\cite{pan}. Thus our work confirms the fact that fundamental concepts about the world can indeed be put to
experimental test.

When this work was completed, we learned from J.-A. Larsson that he has come to similar conclusions using a
somewhat related approach \cite{larsson}. C. S. would like to thank L. Hardy for a useful discussion. This work has
been supported by the Austrian Science Foundation (FWF), projects S6504 and F1506, and by the QIPC Program of the
European Union.


\begin{references}
\bibitem{bell} J. S. Bell, Physics (Long Island City, N.Y.) 1, 195
(1964).

\bibitem{ks} S. Kochen and E. P. Specker, J. Math. and Mech. {\bf 17}, 59
(1967).

\bibitem{meyer} D. Meyer, Phys. Rev. Lett. {\bf 83}, 3751 (1999).
\bibitem{ksexp} A. Cabello and G. Garc\'{\i}a-Alcaine, Phys. Rev. Lett. {\bf 80}, 1797 (1998);
C. Simon, M. \.{Z}ukowski, H. Weinfurter, and A. Zeilinger, Phys. Rev. Lett. {\bf 85}, 1783 (2000); M. Michler, H.
Weinfurter, and M. \.{Z}ukowski, Phys. Rev. Lett. {\bf 84}, 5457 (2000).
\bibitem{clauser} See e.g. J. F. Clauser and M. A. Horne, Phys. Rev. D {\bf
10}, 526 (1974). J. F. Clauser and A. Shimony, Rep. Prog. Phys. {\bf 41}, 1881 (1978).
\bibitem{kent} Meyer's construction was generalized by A. Kent, Phys. Rev. Lett. {\bf 83}, 3755 (1999).

\bibitem{samples}  The principle of experimental tests of classes of hidden-variable
theories is always similar: one verifies that some properties $P_1, P_2,
... P_{n-1}$ hold for almost all systems coming from a certain
source by verifying that they hold for sufficiently large random samples. Based on this, one can make a prediction
for some property $P_n$, which is then again tested on a random sample. This assumes that it is possible for the
observer to obtain representative samples. This means that there should be no correlation between the hidden
variables of the system under study and the observer's choice which observable to measure. For an experiment where
a physical random number generator was used in order to determine the choice of observable to be measured, see G.
Weihs, T. Jennewein, C. Simon, H. Weinfurter, and A. Zeilinger, Phys. Rev. Lett. {\bf 81}, 5039 (1998). It is never
possible to exclude ``conspiracy-type'' hidden-variable theories where the decision of the observer which
observable to measure is determined by the hidden variables (of the world) in such a way that it is impossible for
him to obtain representative samples.

\bibitem{stochastic} In the present paper we do not discuss
stochastic hidden variable theories explicitly. This does not limit the generality of the results derived because
the existence of a stochastic local or non-contextual hidden variable model for a given physical system implies
that also an underlying deterministic local or non-contextual model can be constructed which reproduces the
probabilities of the stochastic model. Therefore e.g. ruling out all possible non-contextual deterministic
hidden-variable models implies ruling out all possible non-contextual stochastic models as well. A stochastic
hidden-variable theory is non-contextual if for every possible value of the hidden variables the probability to
find a certain result for a given observable does not depend on which other observables are measured jointly, and
joint probabilities for several observables are simply obtained by multiplication of the probabilities for single
observables. Note that because of this last requirement quantum mechanics is not a non-contextual
``hidden-variable'' theory.

\bibitem{context} In the ideal case one could define
non-contextuality in such a way that the predetermined value of some quantum mechanical observable $X$ is required
to be independent of the simultaneously measured observables only if they exactly commute with $X$, i.e. only for
ideal Von Neumann measurements. In practice observables corresponding to precise directions do not have an
operational meaning. It is evident that this weaker form of non-contextuality cannot be tested experimentally.

\bibitem{clifton} Let us note that the models discussed by R.
Clifton and A. Kent, quant-ph/9908031, are not non-contextual in the present sense because in these models the
result of a measurement of $S^2_{\vec{n}}$ in general does not only depend on $\lambda$, $\mu$, and $\vec{n}$, but
also on the other observables measured simultaneously.

\bibitem{peres} A. Peres, J. Phys. A {\bf 24}, L175 (1991), showed
how a contradiction can be demonstrated using 20 triads.
\bibitem{mermin2} N. D. Mermin, quant-ph/9912081.
\bibitem{apple} D. M. Appleby, quant-ph/0005056; D. M. Appleby, quant-ph/0005010;
H. Havlicek, G. Krenn, J. Summhammer, and K. Svozil, quant-ph/9911040.
\bibitem{ghz} D. M. Greenberger, M. Horne, A. Shimony, and A. Zeilinger, Am. J.
Phys. {\bf 58}, 1131 (1990). For inequalities applicable to the GHZ situation see N. D. Mermin, Phys. Rev. Lett.
{\bf 65}, 1838 (1990).
\bibitem{pan} J.W. Pan, D. Bouwmeester, M. Daniell, H. Weinfurter, and A.
Zeilinger, Nature {\bf 403}, 515 (2000).
\bibitem{larsson} J.-A. Larsson, quant-ph/0006134.


\end{references}
\end{document}